\documentclass[prl,reprint,superscriptaddress,showpacs,aps]{revtex4-1}

\usepackage{graphicx}

\begin{document}

\title{Short-depth trial-wavefunctions for the variational quantum eigensolver based on the problem Hamiltonian}

\author{Gian Salis}
\author{Nikolaj Moll}

\affiliation{IBM Research -- Zurich, S\"aumerstrasse 4, 8803 R\"uschlikon, Switzerland}

\date{\today}

\begin{abstract}
For the variational quantum eigensolver we propose to generate trial wavefunctions from a small amount of selected Pauli terms of the problem Hamiltonian. Two different approaches, one inspired by the quantum approximate optimization algorithm and the other by imaginary-time evolution, are proposed and studied in detail. Using numerical calculations, we study the efficiency of these trial wavefunctions for finding the ground-state energy of three molecules: H$_2$, LiH and H$_2$O. We find that only a small number of Pauli terms are needed to reach chemical accuracy, leading to short-depth quantum circuits with a small number of variational parameters. For the LiH molecule, the quantum circuit consists of 36 two-qubit gates, 45 one-qubit gates, and four variational parameters, with a favorable scaling for larger molecules.
\end{abstract}


\maketitle

There are only a few applications for which algorithms are known that run exponentially faster for quantum computers. The simulation of quantum systems is one of them~\cite{feynman_simulating_1982, abrams_quantum_1999}. In particular, quantum computers could revolutionize the simulation of molecules, materials or other model Hamiltonians~\cite{aspuru-guzik_simulated_2005, lanyon_towards_2010, whitfield_simulation_2011, omalley_scalable_2016, reiher_elucidating_2017, colless_computation_2018}. Typically, the interest lies in the ground state of the Hamiltonian, from which multitudes of other properties can be calculated. One way to determine the ground-state energy is by using the phase estimation algorithm in which the Hamiltonian of the quantum system is evolved in time on a quantum computer~\cite{aspuru-guzik_simulated_2005, wecker_gate-count_2014}. For noisy intermediate-scale quantum (NISQ) devices, the variational quantum eigensolver (VQE)~\cite{peruzzo_variational_2014, mcclean_theory_2016, moll_quantum_2018, li_variational_2019, parrish_quantum_2019} is especially promising: there, a part of the computational load is transferred from the quantum computer to the classical computer. On the quantum computer, trial wavefunctions that depend on some classical parameters are generated and its energy is determined by measuring suitable expectation values. The classical computer then varies the classical parameters (i.e.\ the trial wavefunction) with the goal of minimizing the energy. A small number of variational parameters is preferred to make it easier to reach the minimum.

In contrast to quantum chemistry on classical computers where the exponentially large Hilbert space makes it impossible to store the full wavefunction of large molecules, on a quantum computer the Hilbert space is naturally provided by the quantum state of the qubits. One of the main challenges of VQE is to find a good parametrization of the trial wavefunction such that states in the Hilbert space can be accessed that are as close as possible to the unknown ground state of the problem Hamiltonian. In addition, to handle the classical optimization part of the VQE algorithm, the number of parameters should be as small as possible, i.e. not scale exponentially with the problem size. Heuristic \cite{kandala_hardware-efficient_2017}, unitary coupled-cluster \cite{barkoutsos_quantum_2018, romero_strategies_2018}, quantum approximate optimization algorithm (QAOA) \cite{farhi_quantum_2014, farhi_quantum_2016} and other parametrizations have been proposed and successfully applied~\cite{dallaire-demers_low-depth_2018}. In VQE, the preparation of the trial wavefunction takes up most of the quantum circuit, and  there is currently a large interest to implement VQE on NISQ devices with a small number of quantum gates. To achieve that goal, there are on one side approaches that are oriented towards the available gates of the given quantum hardware \cite{kandala_hardware-efficient_2017,ganzhorn_gate-efficient_2019}. On the other side, the circuit can be composed from selecting excitations based on physical considerations. Most promising is to start with a coupled-cluster trial-wavefunction which consists of cluster terms of single and double excitation operators. On a classical computer a coupled-cluster method has been demonstrated that automatically adapts to any state of an electronic system and converges to the full CI limit \cite{lyakh_adaptive_2010}. The adaptivity is accomplished through a guided selection of a compact set of cluster terms as required for a proper description of the electronic system under consideration. Such adaptive method also can be used for VQE on a quantum computer \cite{grimsley_adaptive_2019}. Independently it has been shown that not all cluster terms are needed for generating the ground state on a quantum computer \cite{hempel_quantum_2018,ryabinkin_qubit_2018,lee_generalized_2019,nam_ground-state_2019}.

In this letter, we propose to parametrize trial wavefunctions by using gates that are based directly on selected terms of the problem Hamiltonian. We discuss two different methods that are inspired by (i) QAOA, and (ii) imaginary time evolution of the problem Hamiltonian. Both can be implemented in the form of a VQE algorithm. We find that accurate solutions to quantum chemistry problems can be found by selecting only a small number of Pauli terms from the full Hamiltonian, thus keeping the quantum circuit at a short depth and thereby making it accessible to NISQ devices. We thereby achieve a trade-off between the number of parameters and the circuit depth - VQE typically fights with a large number of parameters, whereas phase estimation or QAOA require no or few parameter but have an excessively long circuit depth. We show that by increasing the number of selected Pauli terms, the approximated solution can be tuned arbitrarily close to the exact solution. For LiH, for example, the chemical accuracy of 1.6~mH can be obtained with only four Pauli terms (out of 276) and four variational parameters, requiring 36 two-qubit gates and 45 one-qubit gates for the preparation of the trial wavefunction. In order to find the relevant Pauli terms, a search algorithm is proposed that requires multiple calls of the short-depth quantum circuit, thereby moving load to the classical computer. This search increases the computational load only polynomially. In contrast to a heuristic generation of the trial wavefunction, a Hamiltonian-based generation automatically preserves the particle-number of the trial wavefunction. In second quantization the Hamiltonian only consists of pairs and double pairs of creation and annihilation operators.

Generally, any problem Hamiltonian on a quantum computer with $N$ qubits can be written as a sum of $M$ Pauli terms
\begin{equation}
H = \sum_j^M h_j H_j \, ,
\end{equation}
where $h_j$ are complex coefficients and $H_j$ are Pauli terms $H_j = q_1 \otimes ... \otimes q_N$ composed of $N$ single-qubit Pauli operators (including the identity operator $I$), $q_j \in \{I, X, Y, Z\}$. In the worst case, $M$ scales with the power of four in the number of molecular orbitals, but in practice drastically less terms need to be considered. We are interested in finding the energy $E_0$ of the ground state $|\psi_0\rangle$ of the Hamiltonian $H$. In the VQE, a quantum circuit is created that depends on classical parameters $\theta$ and that generates a trial wavefunction $|\psi(\theta)\rangle$. The expectation value of the energy $\langle \psi(\theta) | H | \psi(\theta) \rangle$ is then measured on the quantum computer. A classical computer then varies $\theta$ in order to minimize the energy and thereby find the ground state energy $E_0 = \min_\theta \langle \psi(\theta) | H | \psi(\theta) \rangle$ of the problem~\cite{peruzzo_variational_2014, mcclean_theory_2016, moll_quantum_2018}. 

Exemplarily, we study the Hamiltonian-based trial wavefunctions for three small molecules: H$_2$, LiH, and H$_2$O. More generally, these kinds of trial wavefunctions can be employed to any kind of problem Hamiltonian. For the three studied molecules, their Hamiltonian is represented in second quantization as a sum of one and two-body terms. The minimal STO-3G basis set is used~\cite{whitfield_simulation_2011, kandala_hardware-efficient_2017}. The Hamiltonians are mapped using Qiskit~\cite{akhalwaya_qiskit:_2019} to the qubit space using a fermion-to-qubit transformation, the parity mapping~\cite{bravyi_tapering_2017}, and applying the available two-qubit reduction. For both LiH and H$_2$O the 1s electrons of the Li and O atom are frozen. This results in a mapping of the problem to two qubits and a Hamiltonian of four Pauli terms for H$_2$, eight qubits and 276 Pauli terms for LiH, and 10 qubits and 551 Pauli terms for H$_2$O. Although the number of Pauli terms $M$ only increases polynomially with the number of qubits $N$~\cite{aspuru-guzik_simulated_2005}, for LiH and H$_2$O the number is already too large to create Hamiltonian-based trial wavefunctions on NISQ devices with limited coherence \cite{kandala_hardware-efficient_2017}. However, we show that only a subset $S \subset \{1,...,M\}$ of $K$ Pauli terms of the total $M$ Pauli terms are needed to find an accurate solution, and we describe how to select those $K$ terms. 

We first discuss how to construct a trial wavefunction based on the idea of QAOA. In standard QAOA, the eigenstate of a drive Hamiltonian is evolved in discrete steps to the ground state of the problem Hamiltonian. The ground state is thereby obtained directly, i.e.\ without any variational parameter, but the implementation of the full Hamiltonian is expensive as it consists of a very large sum of Pauli terms. Here, we parametrize the drive and the problem Hamiltonian and use the evolved wavefunction for VQE. In our mapping of the molecular wavefunction to the qubit space, the Hartree-Fock state is an eigenstate of the qubit basis (i.e.\ of the Pauli $Z$ terms) and therefore simple to prepare on the quantum computer. The Hartree-Fock state is also already close to the ground state, which is helpful for a fast convergence \cite{sugisaki_open_2019}. This motivates us to use $H_z = \sum_q \beta_q Z_q$ as a drive Hamiltonian and inspired by QAOA mix the drive and the problem Hamiltonian to obtain 

\begin{equation}
| \psi(\gamma, \beta) \rangle =  \prod_l^P \prod_{j \in S} e^{-i \gamma_{l j} H_j} \prod_q^N e^{-i \beta_{l j q} Z_q}  | \psi_{\rm start} \rangle \, ,
\label{eq:qaoa}
\end{equation}
where $H_j$ are the selected Pauli terms of the problem Hamiltonian, $Z_q$ are the Pauli operators for the drive Hamiltonian, $P$ the discretization steps, $N$ the number of qubits and $\gamma_{l j}$ and $\beta_{l j q}$ the variational parameters. The starting wavefunction $| \psi_{\rm start} \rangle$ is the Hartree-Fock wavefunction of the molecule. 

We study our QAOA-inspired trial wavefunction for the three molecules as a function of the number of Pauli terms $K$. We start with H$_2$ and select one Pauli term from the four that describe the problem Hamiltonian. For this selected Pauli term we calculate the energy minimum by applying the VQE algorithm. For a discretization step $P = 1$ we obtain the exact solution with the Pauli term {\it XX}: 
\begin{equation}
| \psi(\gamma, \beta) \rangle = e^{-i \gamma XX} e^{-i \beta_1 IZ} e^{-i \beta_2 ZI} | \psi_{\rm start} \rangle \, ,
\end{equation}
with $\gamma = -0.1118$, $\beta_1 = 0.5448$, and $\beta_2 = -0.2406$. The Pauli term {\it XX} together with the $Z$ terms of the drive Hamiltonian swap an electron from the ground state to the excited state. The state $| \psi(\gamma, \beta) \rangle$ completely overlaps with the ground-state subspace for all bond lengths.

\begin{figure}[tb]
\begin{center}
\includegraphics[width=\linewidth]{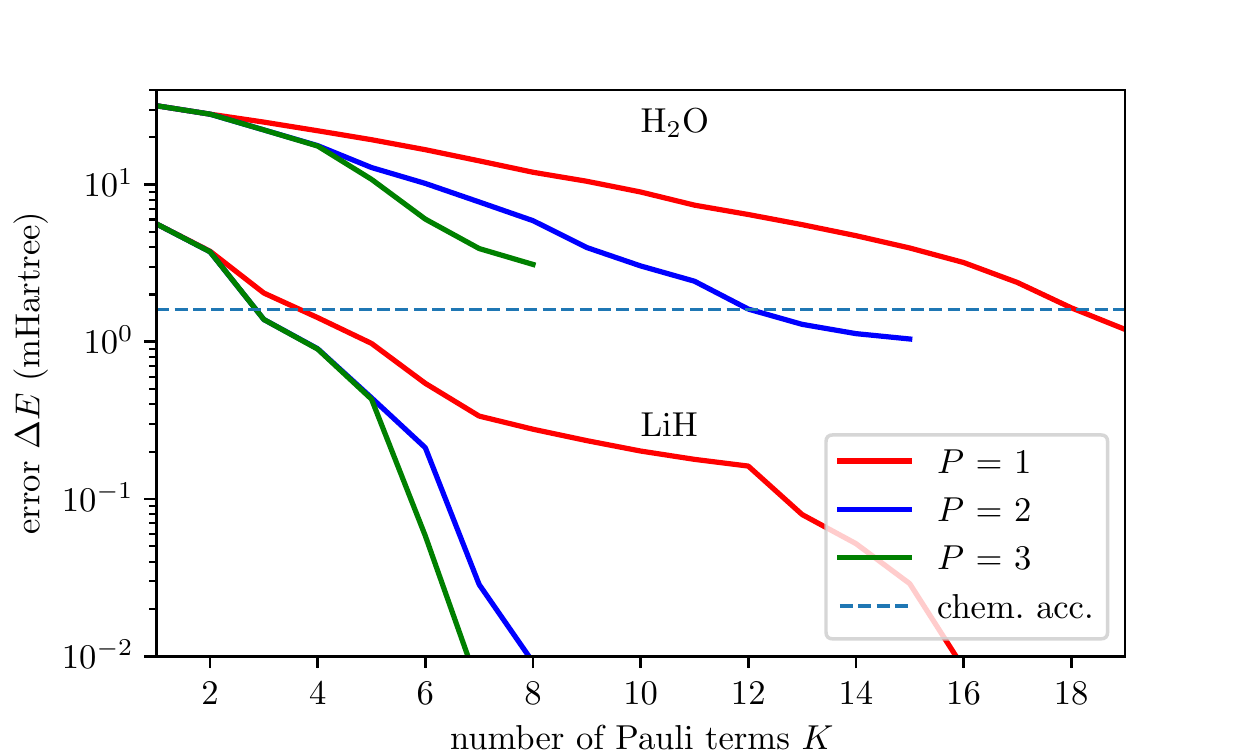}
\caption{\label{fig1} The calculated energy difference of the QAOA solution and the exact solution as a function of the number of Pauli terms $K$ for different number of discretization steps $P$.}
\end{center}
\end{figure}

For the two other molecules, LiH and H$_2$O, the energy difference to the exact solution is calculated for the equilibrium bond length and is shown in Fig.\ \ref{fig1}.
We start with $P=1$ and $K=1$. We randomly select one Pauli term $H_j$ with $j\in\{1,...,M\}$ and apply the algorithm (i.e. generate the trial wave function and minimize the energy by varying the parameter(s)). We then search for the Pauli term where the overall lowest energy is achieved. After finding such a first Pauli term, we search for a second term that together with the selected first term decreases the energy difference even further. To obtain chemical accuracy, the optimal terms are thereby added sequentially. We find that four Pauli terms are needed for LiH and 18 for H$_2$O. 

The results are shown in Tab.\ \ref{tab1} for comparison.
\begin{table}[tb] 
\caption{\label{tab1} Number of qubits $N$ and total number of Pauli terms $M$ needed to represent the three molecules H$_2$, LiH and H$_2$O. Number of optimal Pauli terms $K$ to obtain chemical accuracy, variational parameters, one-qubit gates and two-qubit gates for QAOA-inspired trail wavefunction, for the imaginary time wavefunction and for the UCC trial-wavefunction.}
\begin{tabular}{llrrrrr}
\hline\hline
molecule & & H$_2$ & LiH & H$_2$0 \\ \hline
\multicolumn{2}{l}{qubits $N$}            & 2 &   8 &  10 & $N$\\
\multicolumn{2}{l}{total Pauli terms $M$} & 5 & 276 & 551 & $M < N^4$\\\hline
Pauli terms $K$  & QAOA-inspired  &  1 &   4 &  18 & $K \ll N^4$ \\
                 & imaginary time &  1 &   4 &  18 & $K \ll N^4$\\ \hline
var.\ parameters & QAOA-inspired  &  3 &  36 & 198 & $(N+1)K P$\\
                 & imaginary time &  1 &   4 &  18 & $K P$\\
                 & UCC            &  1 &   8 &  28 & \\ \hline
one-qubit gates  & QAOA-inspired  &  7 &  80 & 432 & $(3 N + 1) K P$\\
                 & imaginary time &  5 &  45 & 302 & $(2 N + 1) K P$ \\ 
                 & UCC            &  5 & 120 & 532 & \\ \hline
two-qubit gates  & QAOA-inspired  &  2 &  36 & 240 & $2 (N - 1) K P$\\
                 & imaginary time &  2 &  36 & 248 & $2 (N - 1) K P$\\
                 & UCC            &  2 & 112 & 504 & \\
\hline\hline
\end{tabular}
\end{table}
Therefore, trial wavefunctions that reach chemical accuracy can be constructed by using only a subset of the full list of terms of the Hamiltonian $H$. The coefficients of the terms in the subset are not given by the Hamiltonian anymore but need to be varied, thereby compensating the error from neglecting a large number of Pauli terms. Even less Pauli terms are needed when increasing the discretization step $P$. This helps because it introduces more variational parameters. It opens up the solution space because the same non-commuting Pauli terms $H_j$ are repeatedly applied, similar to a Trotter expansion. For two discretization steps, three Pauli terms are sufficient for LiH and 12 for H$_2$O. For three discretization steps, 10 Pauli terms (out of 551 terms of the full Hamiltonian) suffice to reach chemical accuracy for H$_2$O.

As stated above, we keep the previously selected Pauli terms when adding a new one. The number of iterations to find the $K$ best Pauli terms out of $M$ scales therefore only polynomially with $M$. If we instead tried to find the $K$ best Pauli terms at once, the number of iterations would scale exponentially with $M$ or more precisely like $M \choose K$.  

\begin{figure}[tb]
\begin{center}
\includegraphics[width=\linewidth]{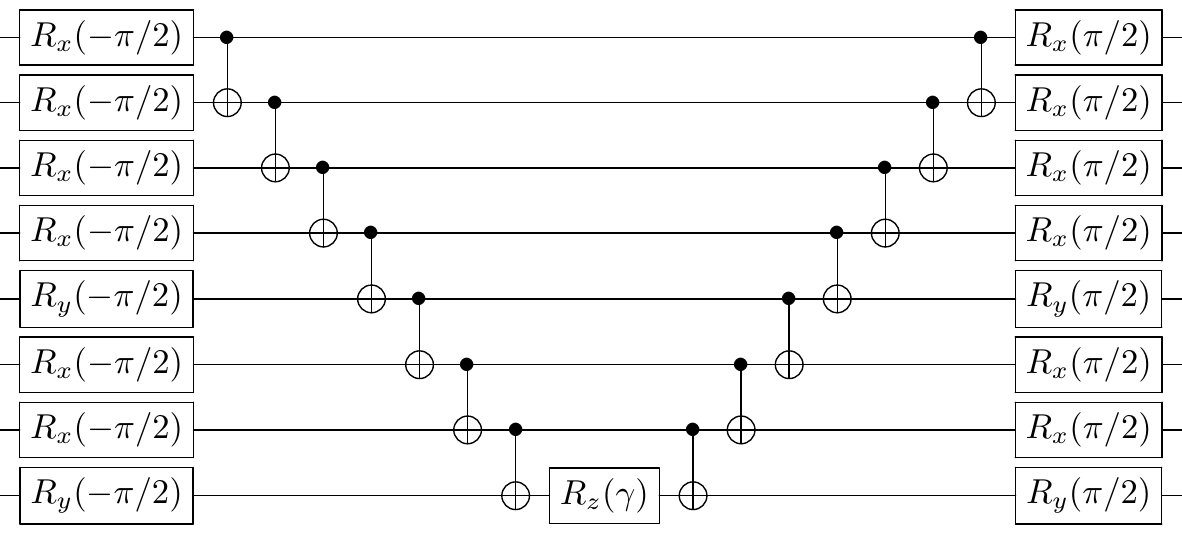}
\caption{\label{fig2} The quantum circuit to implement the term in Eq.~\ref{eq:U} (the first term for LiH) on a quantum device. In total, 17 one-qubit gates and 14 two-qubit gates are necessary.}
\end{center}
\end{figure}

The implementation of a Pauli term on a quantum device needs to apply a unitary operator that consists of the exponentiation of the complex Pauli term. As an example, the Pauli term  {\it YXXYXXXX} results in the unitary operator
\begin{equation}
U = e^{-i \gamma YXXYXXXX} \, .
\label{eq:U}
\end{equation}
The corresponding quantum circuit \cite{barkoutsos_quantum_2018} is shown in Fig.\ \ref{fig2}.
The quantum circuit to create our QAOA-inspired trial wavefunction consists at most of $(3 N + 1) K P$ one-qubit gates and $2 (N - 1) K P$ two-qubit gates. The number of gates reduces when there are identity operators in the Pauli terms. The four Pauli terms needed for LiH to obtain chemical accuracy with one discretion step result in only 80 one-qubit gates and 36 two-qubit gates. The corresponding number of variational parameters is 36 (32 different single-qubit coefficients $\gamma$ and the four two-qubit coefficients $\beta$). The number of gates of the quantum circuits can be reduced by compilation or implementing more complex gates in hardware \cite{ganzhorn_gate-efficient_2019}. The $Z$ terms are typically implemented not as gates on the quantum device but in software by adjusting the phases of the individual qubits. The numbers of gates are smaller than for the unitary coupled-cluster (UCC) trial-wavefunctions, see Tab.\ \ref{tab1}. However, also for the UCC trial-wavefunctions the number of gates can be reduced by selecting the relevant terms only \cite{grimsley_adaptive_2019,nam_ground-state_2019}.

Next we discuss the construction of a Hamiltonian-based trial wavefunction that requires much less variational parameters. It is inspired from the notion that the imaginary-time evolution of a suitable start wavefunction leads with time $t$ to the ground-state wavefunction
\begin{equation}
\label{imagtime}
| \psi(t) \rangle = e^{-t H} | \psi_{\rm start} \rangle \, .
\end{equation}
The imaginary-time evolution cannot be directly implemented on the quantum computer. The corresponding evolution operator
\begin{equation}
e^{-t H_j} = \cosh t - \sinh t H_j\ 
\end{equation}
is not unitary and does not conserve the norm. To approximate this operator with a unitary operator that can easily be implemented on a quantum computer, we construct a Hermitian operator $H'$ from the Hamiltonian $H$ such that $U' | \psi_{\rm start} \rangle = e^{-i t H'}  | \psi_{\rm start} \rangle$ overlaps with the solution subspace of $e^{-t H} | \psi_{\rm start} \rangle$. 

We exemplify this construction with the hydrogen molecule encoded on two qubits. Numerically solving Eq.~\ref{imagtime}, we find that for H$_2$, the most relevant Pauli term is $XX$. This is the same as for the QAOA-inspired trial wavefunction. The wavefunction after imaginary-time evolution in the $XX$ term is given by 
\begin{eqnarray}
e^{-t XX} |10\rangle &=& \cosh t |10\rangle - \sinh t |01\rangle \nonumber\\
 &=& a_{xx} |10\rangle - b_{xx} |01\rangle\, ,
\end{eqnarray}
which needs to be normalized. A similar superposition can be obtained after real-time evolution of the Hermitian operator $H' = XY$:
\begin{eqnarray}
e^{-i t XY} |10\rangle &=& \cos t |10\rangle - \sin t |01\rangle  \nonumber\\
 &=& a_{xy} |10\rangle - b_{xy} |01\rangle\,\, .
\end{eqnarray} 

\begin{figure}[tb]
\begin{center}
\includegraphics[width=\linewidth]{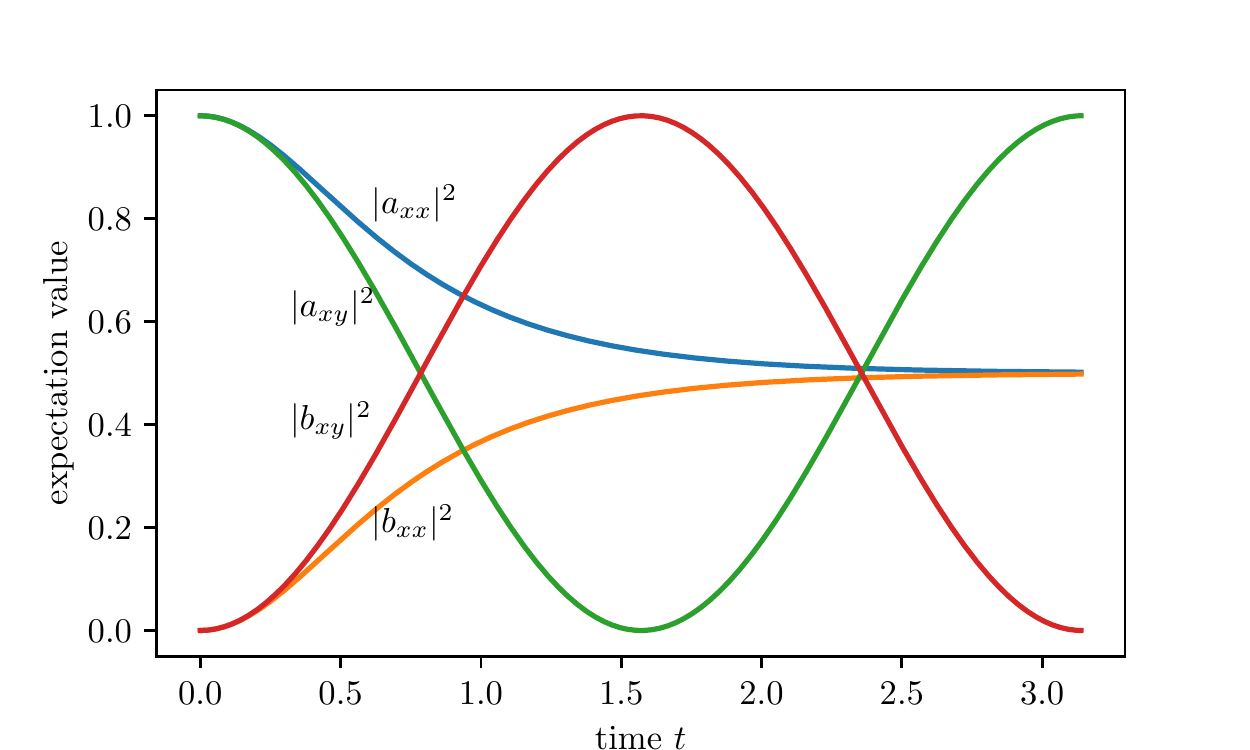}
\caption{\label{fig3} The square of the renormalized amplitudes of the trial states of the hydrogen molecule as a function of the variational parameter $t$. The labels $|a_{xx}|^2$ and $|b_{xx}|^2$ refer to the ground and excited state of the imaginary-time evolution in the $XX$ term. The labels $|a_{xy}|^2$ and $|b_{xy}|^2$ are the respective quantities for a unitary evolution due to the $XY$ term.}
\end{center}
\end{figure}

The squared amplitudes of the state created by the imaginary-time-evolution in the $XX$ term and of the state evolving by the unitary $U' = e^{-i t H'}$ can be seen in Fig.\ \ref{fig3}.
For molecules, the amplitudes are real numbers. The imaginary-time evolution results in an equal superposition of $|01\rangle$ and $|10\rangle$. The constructed unitary operator $U'$ leads to an oscillation of the amplitudes with $t$, reaching the equal superposition at $t=\pi/4$. We can therefore use $t$ as a a variational parameter in $U'$.

From this we derive the following heuristic for the algorithm inspired by the imaginary-time evolution: We replace in each Pauli term $H_j$ one {\it X} with one {\it Y} or vice versa. When using this modified Pauli term $H'_j$, the corresponding $U'$ mimics the imaginary-time evolution of $H_j$. For example, for LiH the optimal Paul term  $H_j = YXXYXXXX$ becomes $H'_j = YYXYXXXX$. It does not matter for which qubit we exchange {\it X} with {\it Y}, it is only important that by this replacement the Pauli term $H_j$ acquires a factor $i$. This is seen by considering that the Pauli matrices for {\it X} and {\it Y} differ by a factor $i$ (besides a minus sign of one off-diagonal element that however does not affect the efficiency of the proposed algorithm). The Pauli term $XX$ of the H$_2$ Hamiltonian becomes this way the excitation operator $XY$ of the coupled cluster method. The trial wavefunction based on imaginary-time evolution is then constructed by
\begin{equation}
| \psi(\gamma) \rangle =  \prod_l^P \prod_{j \in S} e^{-i \gamma_{l j} H'_j} | \psi_{\rm start} \rangle \, .
\end{equation}
Here, each Pauli term $H_j$ has one parameter $\gamma_{l j}$ for each Trotter step $P$. Increasing the number of Trotter steps helps in the same way as increasing the discretization steps in the QAOA-inspired approach. Altogether there are $K P$ parameters in this approach. The term $e^{-i \gamma_{l j} H'_j}$ is unitary and can be implemented on a quantum device, as exemplified in Fig.\ \ref{fig2}.

\begin{figure}[tb]
\begin{center}
\includegraphics[width=\linewidth]{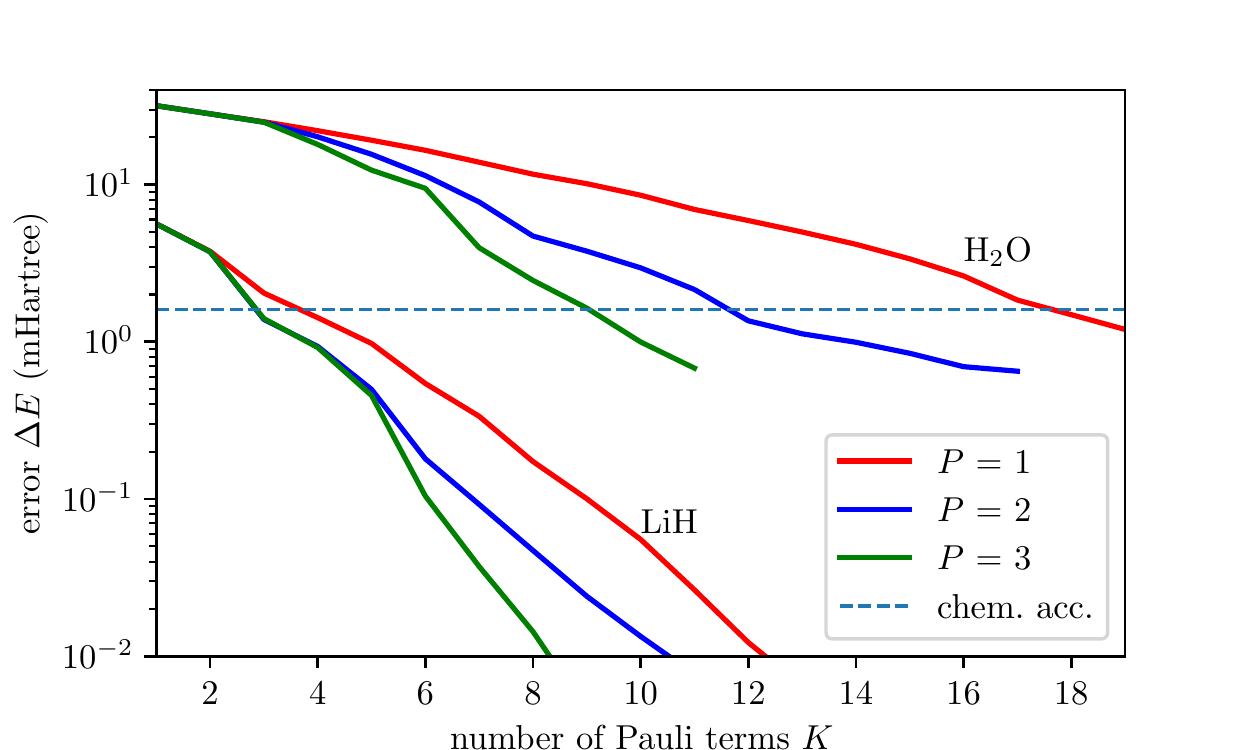}
\caption{\label{fig4} The energy difference between the variational solution based on imaginary-time evolution and the exact solution, plotted as a function of the number of Pauli terms $K$.}
\end{center}
\end{figure}

The energy difference between the imaginary time-evolution solution and the exact solution as a function of the number of Pauli terms $K$ is shown in Fig.\ \ref{fig4} for both LiH and H$_2$O. 
The Pauli terms are selected successively one after each other. The error in the optimized energy decreases in a similar way with the number of Pauli terms $K$ as the QAOA-inspired trial wavefunction shown in Fig.\ \ref{fig1}. As for the QAOA-inspired case, four Pauli terms are needed for LiH and 18 for H$_2$O to obtain chemical accuracy for $P=1$. Importantly, the number of variational parameters is only given by the number of selected Pauli terms and therefore much smaller compared to the QAOA-inspired approach (four instead of 36 for LiH, 18 instead of 198 for H$_2$O). When increasing the number of Trotter steps $P$, these numbers increase, but less Pauli terms are needed: three terms for LiH and 12 for H$_2$O with $P=2$, and only nine terms for H$_2$O with $P=3$. The optimal Pauli terms are listed in the appendix.

We have also studied the influence of the bond length of the molecules on our trial wavefunction. For LiH we found that for all bond lengths we only need four Pauli terms to obtain chemical accuracy. However, for larger bond lengths the system becomes more entangled and the first optimal Pauli term changes for bond lengths larger than 2.4~\AA\ from $H'_j = YYXYXXXX$ to $H'_j = YXXXXXXX$.

The implementation of our trial wavefunction based on imaginary-time evolution in a quantum circuit leads to a shorter circuit depth. A maximum of $(2 N + 1) K P$ one qubit gates and $2 (N - 1) K P$ two-qubit gates are needed. Considering the identities in the Pauli terms, for LiH with $K=1$, only 45 one-qubit gates and 36 two-qubit gates are enough to obtain chemical accuracy with only 4 variational parameters. 

We can compare our results to the UCC trial-wavefunction where for LiH, 8 variational parameters, 120 one-qubit gates and 112 two-qubit gates are needed (considering one Trotter step). In comparison, the QAOA-inspired trial wavefunction reduces the number of gates by a third for the one-qubit gates and by more than a factor of three for the two-qubit gates. With the imaginary-time evolution trial-wavefunction, the variational parameters are reduced by 4, the one-qubit gates by more than a factor of two and the two-qubit gates by more than a factor of three. The data is summarized in Tab.~\ref{tab1}. 

In conclusion, we described two types of particle-number-preserving trial-wavefunctions for VQE which are based on the problem Hamiltonian. By selecting a small number of Pauli terms $K$ of the problem Hamiltonian, one can construct quantum circuits that have a very short depth and that create trial wavefunctions arbitrarily close to the exact solution. The Pauli terms are selected by finding the term that leads to the smallest energy. With the first Pauli term being selected, a second Pauli term is searched for. If needed, this is repeated and further Pauli terms are added until the desired accuracy is reached. We find that the trial wavefunction inspired by imaginary time evolution of the problem Hamiltonian leads to a dramatic reduction of the number of variational parameters and at the same time to a decrease of the circuit depth. For example, chemical accuracy can be reached for LiH with four selected Pauli terms, four variational parameters and only 36 two-qubit and 45 one-qubit gates. Such short-depth circuits will become important for assessing NISQ computers with the goal to calculate the ground-state energy of small molecules, other quantum systems or synthetic problem Hamiltonians. 

We thank Panagiotis~Barkoutsos, Daniel~Egger, Stefan~Filipp, Andreas~Fuhrer, Marc~Ganzhorn, Ivano~Tavernelli, Stefan~Woerner, the quantum teams at IBM Research -- Zurich and the IBM T.~J.~Watson Research Center for insightful discussions.

\bibliography{paper}

\end{document}